\documentclass[secnumarabic,aps,pra,amsfonts,twocolumn,%
showkeys,floatfix%
]{revtex4-1}

\renewcommand{\bibnumfmt}[1]{\mbox{#1}. }
\usepackage{bm,amsfonts,amsthm}
\usepackage{placeins}
\usepackage{longtable}
\usepackage{graphicx}
\newcommand{\gl}[1]{(\ref{#1})}

\begin{document}
\title{Structural Distortion Stabilizing the Antiferromagnetic
  and Semiconducting Ground State of BaMn$_2$As$_2$}
\author{Ekkehard Kr\"uger} 
\affiliation{Institut f\"ur Materialwissenschaft, Materialphysik,
  Universit\"at Stuttgart, D-70569 Stuttgart, Germany}
%
\date{\today}
\begin{abstract}
  We report evidence that the experimentally found
  antiferromagnetic structure as well as the semiconducting ground
  state of BaMn$_2$As$_2$ are caused by optimally-localized Wannier
  states of special symmetry existing at the Fermi level of
  BaMn$_2$As$_2$. In addition, we find that a (small) tetragonal
  distortion of the crystal is required to stabilize the
  antiferromagnetic semiconducting state. To our knowledge, this
  distortion has not yet been established experimentally.
\end{abstract}

\keywords{BaMn2As2, magnetism, small band gap semiconductor, nonadiabatic Heisenberg model, group theory}
\maketitle

\section{Introduction}
\label{sec:introduction}
\begin{figure*}
\begin{minipage}[c]{.48\textwidth}
\includegraphics[width=.9\textwidth,angle=0]{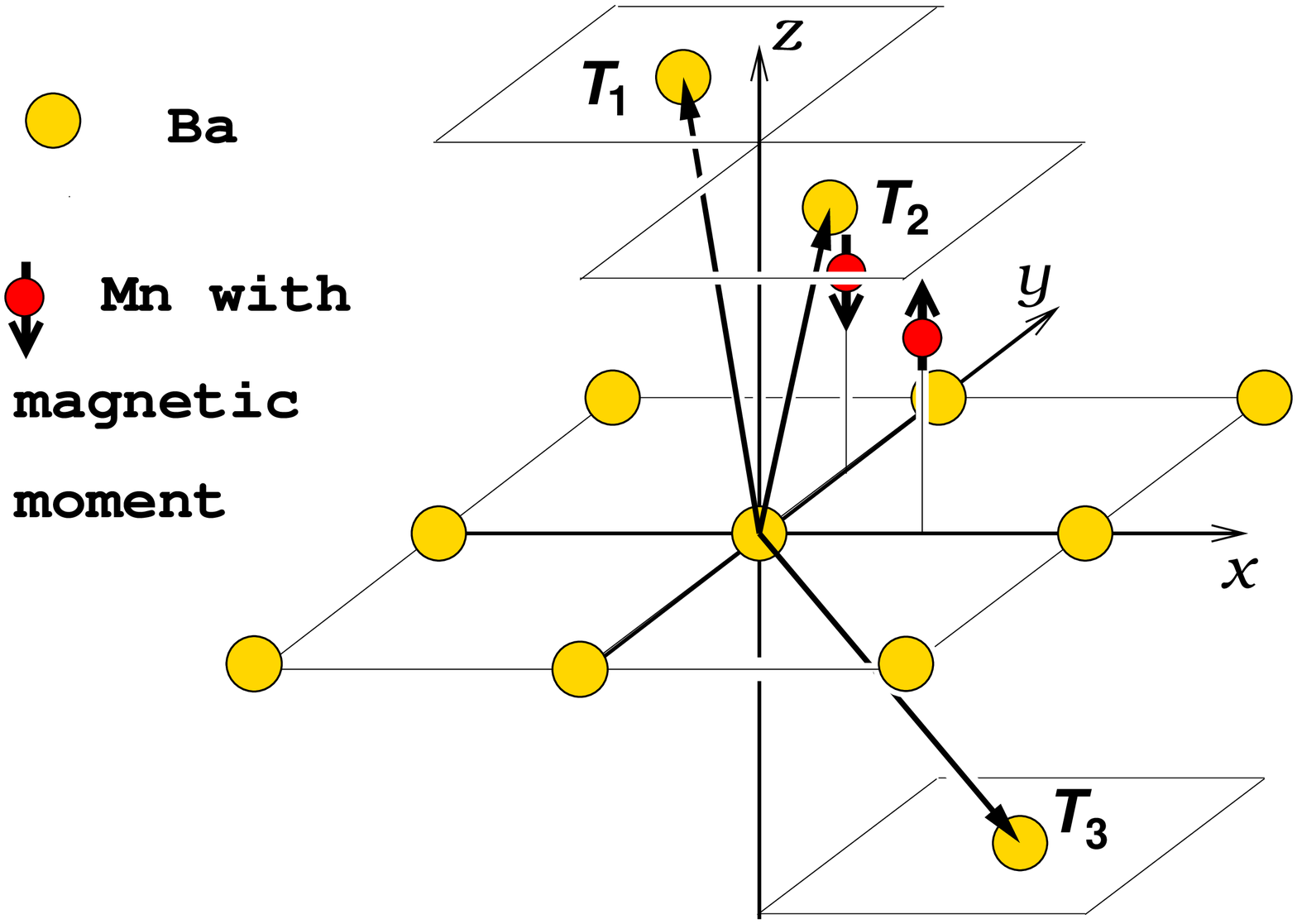}%
\vspace{.8cm}
\begin{center}
(a)\ $I\overline{4}2m = \Gamma^v_qD^{11}_{2d}$ (121)
\end{center}
\end{minipage}
\begin{minipage}{.41\textwidth}
\includegraphics[width=.9\textwidth,angle=0]{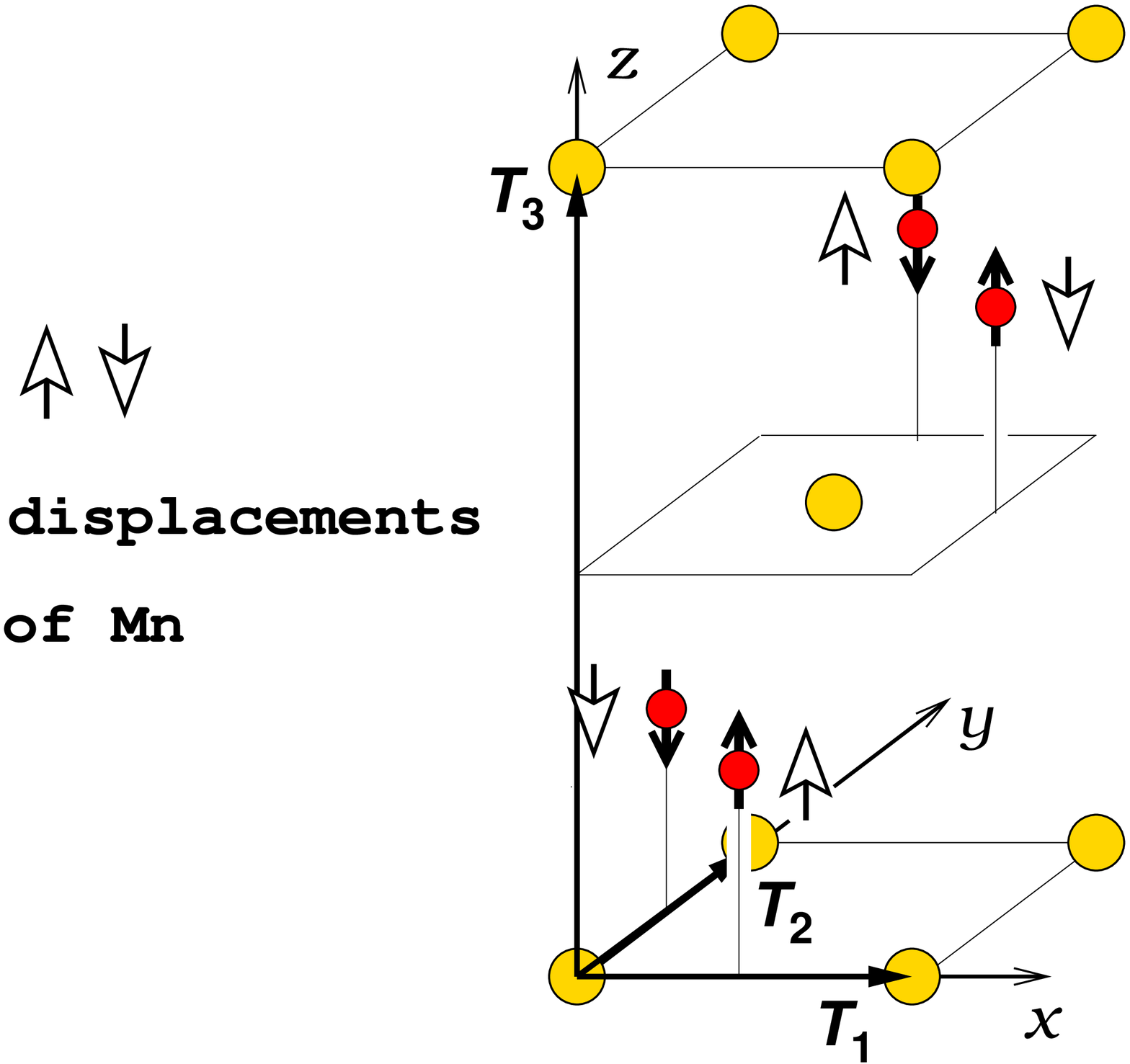}%
\begin{center}
(b)\ $P\overline{4}2_1c = \Gamma_qD^4_{2d}$ (114)
\end{center}
\end{minipage}
\vspace{1cm}
\caption{
Experimentally observed~\cite{singh} antiferromagnetic
structure in  undistorted (a) and distorted (b) BaMn$_2$As$_2$. While sufficient Ba atoms are depicted to
recognize the orientation of the crystal, the Mn atoms are shown only
within the respective unit cell.  The As
atoms are not included. The indicated (small) displacements of the Mn
atoms in exact $\pm \bm T_3$ direction realize the tetragonal
primitive space group $P\overline{4}2_1c$. 
\label{fig:structures}
}
\end{figure*}

The electronic ground state of BaMn$_2$As$_2$ shows resemblances but
also striking differences, as compared with the ground state of the
isostructural compound BaFe$_2$As$_2$. Both materials become
antiferromagnetic below the respective N{\'e}el temperature. However,
while the magnetic moments in BaMn$_2$As$_2$ are orientated along the
tetragonal $c$ axis~\cite{singh}, see Fig.~\ref{fig:structures}, they
are orientated perpendicular to this axis in
BaFe$_2$As$_2$~\cite{huang2}.

Also very interesting is the observation that, unlike BaFe$_2$As$_2$,
BaMn$_2$As$_2$ is a small band gap antiferromagnetic {\em
  semiconductor}~\cite{an,singh2}.  No structural transformation or
distortion of BaMn$_2$As$_2$ in this antiferromagnetic semiconducting
state was experimentally detected~\cite{singh}.

The present paper reports evidence that the remarkable features of the
electronic ground state of BaMn$_2$As$_2$ are connected with
optimally-localized Wannier functions existing at the Fermi level of
BaMn$_2$As$_2$. These Wannier functions are adapted to the symmetry of
the antiferromagnetic structure and constructed from Bloch functions
of well-defined symmetry forming narrow ``magnetic bands'' as defined
in Ref.~\cite{theoriewf} (see Definition 16 $ibidem$).

In Section~\ref{sec:nonexistence} we shall identify the tetragonal
space group $I\overline{4}2m$ (121) as the space group of the
antiferromagnetic structure observed in BaMn$_2$As$_2$ (the number in
parenthesis is the international number) and determine the magnetic
group $M_{121}$ of this structure. We will show that no magnetic band
related to $M_{121}$ exists in the band structure of
BaMn$_2$As$_2$. The situation changes drastically when in
Section~\ref{sec:existence} we shall consider a slightly distorted
crystal. We will define and verify the existence of a magnetic
``super'' band in distorted antiferromagnetic BaMn$_2$As$_2$. This
super band consists of three magnetic bands as defined in
Ref.~\cite{theoriewf} with Wannier functions situated at the Ba, the
Mn, and the As atoms, respectively.

Our group-theoretical results in Section~\ref{sec:magneticbands} will
be physically interpreted in Section~\ref{sec:interpretation}. We
will argue in Section~\ref{sec:distortion} that a small tetragonal
distortion of the crystal is required to stabilize the
antiferromagnetic semiconducting ground state of BaMn$_2$As$_2$. This
distortion alters the space group $I\overline{4}2m =
\Gamma^v_qD^{11}_{2d}$ of the undistorted antiferromagnetic crystal
into the space group $P\overline{4}2_1c = \Gamma_qD^{4}_{2d}$ (which
is still tetragonal) and may be realized by the displacements of the
Mn atoms depicted in Fig.~\ref{fig:structures} (b). These
displacements are evidently so small that they have not yet been
experimentally verified~\cite{singh}.  In Section~\ref{sec:insulating}
we shall show that evidently the magnetic super band is responsible
for the small band gap in the antiferromagnetic semiconducting ground
state, and in Section~\ref{sec:magneticstructure} why the space groups
of the magnetic structures in BaFe$_2$As$_2$ and BaMn$_2$As$_2$ differ
so strikingly.

\subsection{Nonadiabatic Heisenberg model}
\label{sec:nhm}
The existence of magnetic bands in the band structure of
BaMn$_2$As$_2$ is physically interpreted within the nonadiabatic
Heisenberg model (NHM)~\cite{enhm}. The second postulate of the NHM
(Equation (2.19) of~\cite{enhm}) states that in narrow bands
(i.e. in band satisfying Equation (2.13) of~\cite{enhm}) the electrons
may lower their total correlation energy by condensing into an
atomic-like state as it was described by Mott~\cite{mott} and
Hubbard~\cite{hubbard}: the electrons occupy the localized states as
long as possible and perform their band motion by hopping from one
atom to another. Within the NHM, however, the localized states {\em
  are not represented by (hybrid) atomic orbitals but consequently by
  symmetry-adapted optimally-localized Wannier states}. The electrons
are strongly correlated in this atomic-like state, leading to the
consequence that a consistent description of the localized Wannier
states must involve the nonadiabatic motion of the atomic
cores~\cite{enhm}.

Hence, the nonadiabatic localized functions representing these
nonadiabatic Wannier states depend on an additional coordinate
characterizing the motion of the atomic cores. Fortunately, these
mathematically complicated functions need not be explicitly
known. They can be simply managed within the group-theoretical NHM
because they have the same symmetry as the related adiabatic
optimally-localized Wannier functions as defined in
Ref.~\cite{theoriewf}. In this context we speak of ``adiabatic''
Wannier functions if they do not dependent on the nonadiabatic motion
of the atomic cores.

The total correlation energy of the electron system decreases by the
nonadiabatic condensation energy $\Delta E$ defined in Equation (2.20)
of~\cite{enhm} at the condensation into the nonadiabatic atomic-like
state.

\section{Magnetic bands in the band structure of $\text{BaMn}_2\text{As}_2$}
\label{sec:magneticbands}
\subsection{The space group $I\overline{4}2m$ (121) of the
  antiferromagnetic structure in undistorted
  $\text{BaMn}_2\text{As}_2$}
\label{sec:nonexistence}
Removing from the space group $I4/mmm$ of BaMn$_2$As$_2$ all the
symmetry operations not leaving invariant the magnetic moments of the
Mn atoms, we obtain the group $I\overline{4}2m$ (121) as the space group of the
antiferromagnetic structure in undistorted BaMn$_2$As$_2$. Just as
$I4/mmm$, the group $I\overline{4}2m$ has the tetragonal body-centered
Bravais lattice $\Gamma^v_q$.

The group $I\overline{4}2m$ may be defined by the two ``generating
elements''
\begin{equation}
  \label{eq:1}
  \{S^+_{4z}|000\}\text{ and }\{C_{2x}|000\}, 
\end{equation}
see Table 3.7 of Ref.~\cite{bc}. Just as in all our papers, we write
the symmetry operations $\{R|pqr\}$ in the Seitz notation detailed in
the textbook of Bradley and Cracknell \cite{bc}: $R$ stands for a
point group operation (as defined, e.g., in Table 1.4 $ibidem$) and
$pqr$ denotes the subsequent translation $\bm t = p\bm T_1 + q\bm T_2
+ r\bm T_3$, where the $\bm T_1, \bm T_2, \text {and } \bm T_3$ denote
the basic vectors of the respective Bravais lattice given in
Fig.~\ref{fig:structures}. A (magnetic) structure is invariant under a
space group $G$ if it is already invariant under the generating
elements of $G$. The two generating symmetry operations \gl{eq:1}
leave invariant the atoms of BaMn$_2$As$_2$ since both operations are
elements of $I4/mmm$. By means of Fig.~\ref{fig:structures} (a) we can
realize that they additionally leave invariant the magnetic structure,
cf. Sec. 3.1 of Ref.~\cite{lafeaso1}.

 \begin{figure*}[!]
 \includegraphics[width=.85\textwidth,angle=0]{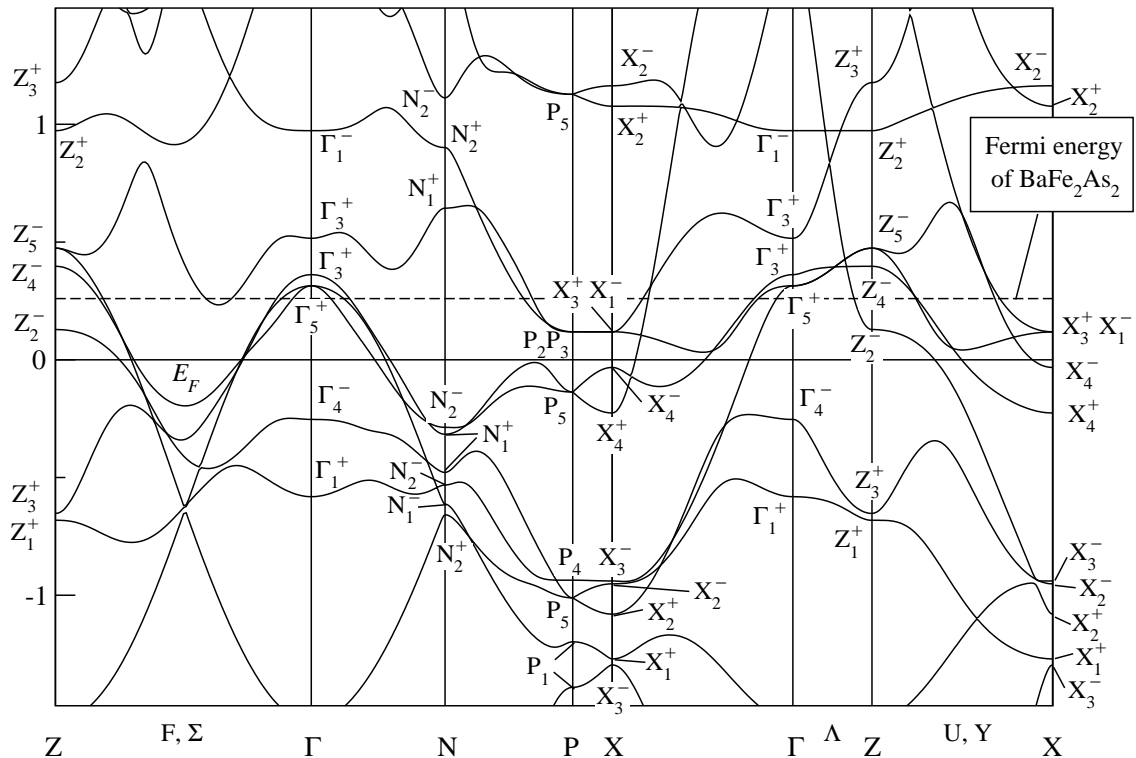}%
 \caption{ Band structure of BaMn$_2$As$_2$ as calculated by the FHI-aims program
\cite{blum1,blum2}, using the structure parameters given in
Ref.~\cite{singh}. The space group of BaMn$_2$As$_2$ is the tetragonal 
group $I4/mmm$ (139)~\cite{singh}, the given symmetry labels are determined by the author. The
notations of the points and lines of symmetry in the Brillouin zone
for $\Gamma^v_q$ follow Fig. 3.10 (b) of Ref.~\cite{bc}, and the
symmetry labels are defined in Table~2 of Ref.~\cite{bafe2as2}. $E_F$
denotes the Fermi level. The
band structure of BaMn$_2$As$_2$ essentially concides with the band
structure of BaFe$_2$As$_2$ (depicted in Fig. 2 of
Ref.~\cite{bafe2as2}) when the Fermi level is moved upwards to the
dashed line.
 \label{fig:bandstr139}
}
 \end{figure*}


 \begin{figure*}[!]
 \includegraphics[width=.85\textwidth,angle=0]{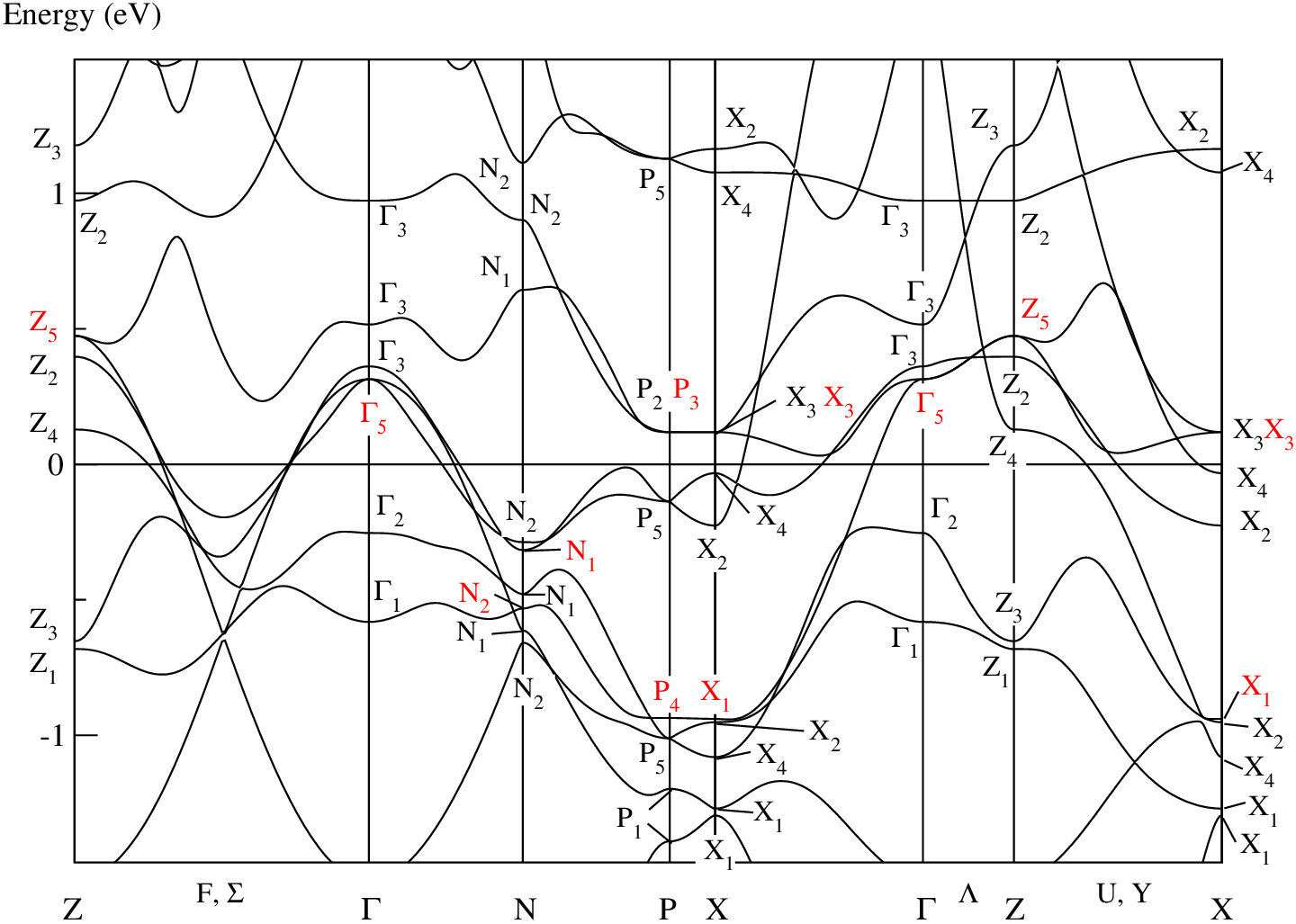}%
 \caption{Band structure of BaMn$_2$As$_2$ as given in Fig.\
 \ref{fig:bandstr139} with symmetry labels of the space group 
 $I\overline{4}2m$ (121) of the antiferromagnetic structure in
 undistorted BaMn$_2$As$_2$. The symmetry labels are determined from
 Tab.\ \ref{tab:falten139_121} and the red labels define band 2 of Mn in 
 Tab.\ \ref{tab:wf121}.
 \label{fig:bandstr121}
}
 \end{figure*}


The associated magnetic group reads as
\begin{equation}
  \label{eq:2}
  M_{121} = I\overline{4}2m + \{KI|000\}I\overline{4}2m,
\end{equation}
where $K$ and $I$ denote the operator of time inversion and the
inversion, respectively. $\{KI|000\}$ leaves invariant both the atoms
and the magnetic structure since $\{I|000\} \in I4/mmm$.

With consideration of the change of symmetry by the magnetostriction,
but neglecting all other magnetic interactions, we receive from the
band structure of BaMn$_2$As$_2$ given in Fig.~\ref{fig:bandstr139}
the band structure of antiferromagnetic undistorted BaMn$_2$As$_2$
depicted in Fig.~\ref{fig:bandstr121}. All the possible magnetic bands
(Definition 16 of Ref.~\cite{theoriewf}) in the magnetic group
$M_{121}$ are listed in Table~\ref{tab:wf121}. The ``best'' magnetic
band would be band 2 of Mn as highlighted in Fig.~\ref{fig:bandstr121}
by the red labels.
 
Band 2 of Mn, however, is not a magnetic band in BaMn$_2$As$_2$
because it misrepresents the Bloch functions at parts of the Fermi
level. Between the $N_1$ and $P_3$  states it jumps over the
Fermi level simulating in this way Bloch states at the Fermi level
which do not exist. The same situation we have between the $Z_5$ state
and the two $X_3, X_1$ states. Along the lines F, $\Sigma$ the
$\Gamma_5$ state is connected with two Bloch states at the Fermi
level, which, however, are not connected to $Z_5$.

We could try to render ineffective these unfavorable jumps by adding
further bands to the Mn band as it will be successful in the space
group $P\overline{4}2_1c$ considered in the following
Section~\ref{sec:existence}. By means of Table~\ref{tab:wf121} we may
satisfy ourselves that this procedure is not possible. For instance,
we neither can add band 1 nor band 2 of As to band 2 of Mn because
there is neither a $\Gamma_4$ state nor an additional $P_3$ state
available in die band structure.

\subsection{The space group $P\overline{4}2_1c$ (114) of the
  antiferromagnetic structure in distorted
  $\text{BaMn}_2\text{As}_2$}
\label{sec:existence}
The situation described in the preceding
Section~\ref{sec:nonexistence} changes drastically when we consider
the space group $P\overline{4}2_1c$. This group has no longer the the
tetragonal body-centered Bravais lattice $\Gamma^v_q$, but the
tetragonal primitive lattice $\Gamma_q$ and may be defined by the two
generating elements
\begin{equation}
  \label{eq:3}
  \textstyle
  \{S^+_{4z}|000\}\text{ and }\{C_{2x}|\frac{1}{2}\frac{1}{2}\frac{1}{2}\}, 
\end{equation}
see Table 3.7 of Ref.~\cite{bc} (note that the basis vectors now are
given in Fig.~\ref{fig:structures} (b)). As well as the generating
elements of $I\overline{4}2m$~\gl{eq:1} they leave invariant both the
positions of the atoms and the magnetic structure since the vector
$\bm t = (\frac{1}{2}\frac{1}{2}\frac{1}{2})$ is a lattice vector in
$\Gamma^v_q$. In addition, the generating elements~\gl{eq:3} leave
invariant the displacements of the Mn atoms depicted in
Fig.~\ref{fig:structures} (b). Thus these displacements ``realize''
the space group $P\overline{4}2_1c$ in the sense that the electrons
now move in a potential adapted to the symmetry of the distorted
crystal.  The group $P\overline{4}2_1c$ represents only a small
distortion of the crystal because it is still tetragonal and possesses
the same point group as the space group $I\overline{4}2m$ of
antiferromagnetic undistorted BaMn$_2$As$_2$. It is only the
translation $\bm t = (\frac{1}{2}\frac{1}{2}\frac{1}{2})$ which is no
longer a symmetry operation in $P\overline{4}2_1c$.

At first, the two anti-unitary operations $\{KI|000\}$ and
$\{KI|\frac{1}{2}\frac{1}{2}\frac{1}{2}\}$ may define the magnetic
group of the magnetic structure since both operations leave invariant
the magnetic structure. However, only
$\{KI|\frac{1}{2}\frac{1}{2}\frac{1}{2}\}$ leaves additionally
invariant the displacement of the Mn atoms depicted in
Fig~\ref{fig:structures} (b). These displacements, however, are
required to realize the space group $P\overline{4}2_1c$. Hence, the
magnetic group of antiferromagnetic distorted BaMn$_2$As$_2$ may be
written as
\begin{equation}
  \label{eq:4}
  \textstyle
  M_{114} = P\overline{4}2_1c + \{KI|\frac{1}{2}\frac{1}{2}\frac{1}{2}\}P\overline{4}2_1c.
\end{equation}

 \begin{figure*}[!]
 \includegraphics[width=.95\textwidth,angle=0]{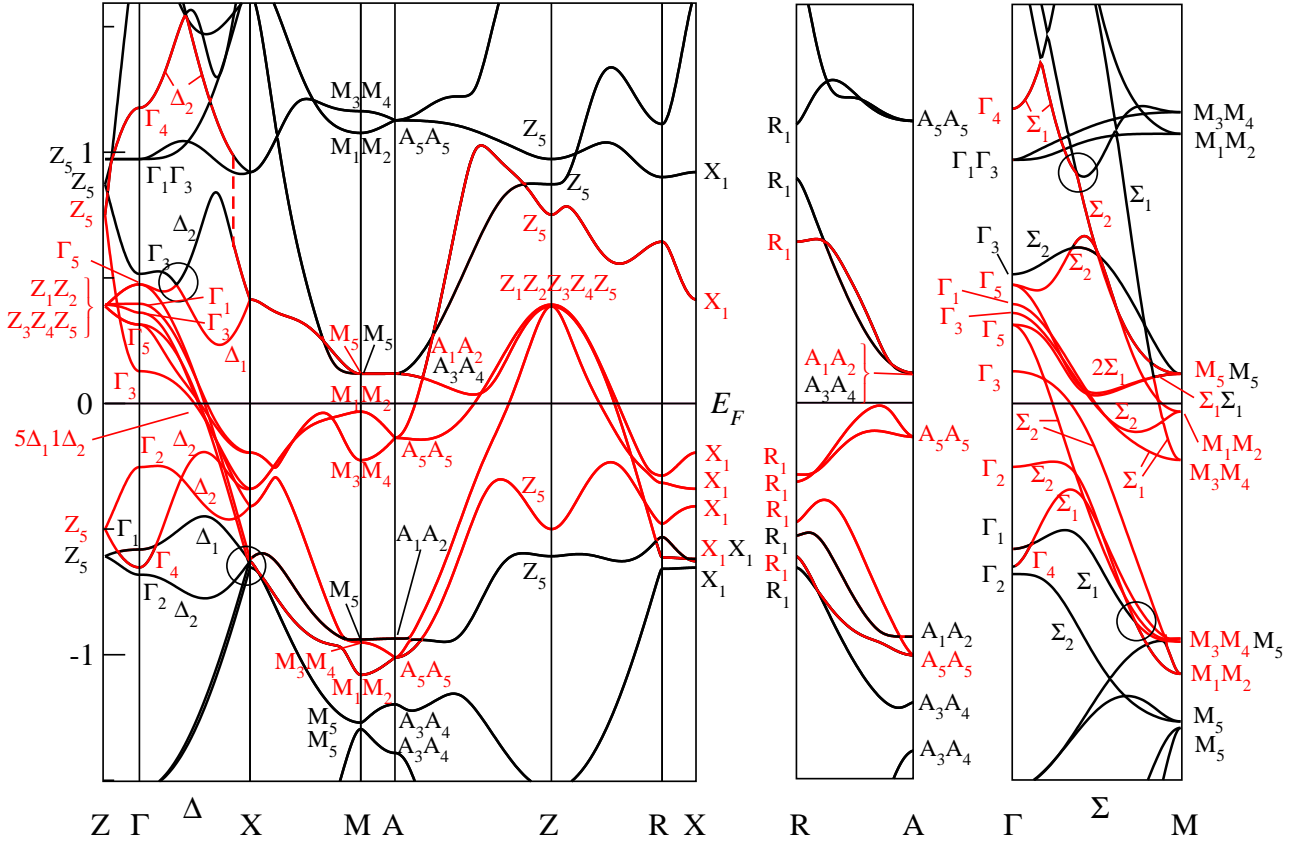}%
\caption{ The band structure of BaMn$_2$As$_2$ as given in Fig.\
 \ref{fig:bandstr139} folded into the
Brillouin zone for the tetragonal primitive Bravais lattice $\Gamma_q$
of the space group $P\overline{4}2_1c$ (114). 
The symmetry labels are defined in
Table~\ref{tab:rep114} and are determined from
Fig.~\ref{fig:bandstr139} by means of Table~\ref{tab:falten139_114}. The notations of
the points of symmetry follow Fig.~3.9~of Ref.~\cite{bc}. $E_F$ denotes the Fermi level. 
The red lines and red
symmetry labels form the
magnetic ``super'' band of the experimentally
observed~\cite{singh} antiferromagnetic structure in
BaMn$_2$As$_2$. It is related to {\em all} the atoms of BaMn$_2$As$_2$, that
means that it is related to two Ba, four Mn and four As atoms in
the unit cell of $\Gamma_q$, and, 
hence, consists of ten branches. Whenever a black and a red line overlap, the red line lies
on the top. The four circles mark the regions with transitions from 
$\Delta_2$ to $\Delta_1$, $\Delta_1$ to $\Delta_2$, $\Sigma_1$ to
$\Sigma_2$, and $\Sigma_2$ to $\Sigma_1$, respectively, on the lines
$\Delta$ and $\Sigma$. These
transitions slightly destroy the symmetry of related the Wannier functions in the
space group $P\overline{4}2_1c$. However, since
these transitions
only occur in two lines of two branches fare away from the Fermi level, we
assume that we may describe the magnetic structure of BaMn$_2$As$_2$ with
high accuracy in the space group $P\overline{4}2_1c$. Nevertheless, these transitions produce
a small additional distortion of the crystal going beyond the distortion depicted
in Fig.~\ref{fig:structures} (b).  
\label{fig:bandstr114}
}
 \end{figure*}


Folding the band structure of BaMn$_2$As$_2$ as given in
Fig.~\ref{fig:bandstr139} into the Brillouin zone for
$P\overline{4}2_1c$ we receive the band structure depicted in
Fig.~\ref{fig:bandstr114}.  All the magnetic bands in the magnetic
group $M_{114}$ are listed in Table~\ref{tab:wf114}. Now we have a very
interesting situation not yet considered in our former papers: We are
able to assign optimally-localized symmetry-adapted Wannier functions
to {\em all the atoms} in the unit cell of $\Gamma_q$, meaning that we
have a band of ten branches with Wannier functions at the two Ba, the
four Mn and the four As atoms. Such a magnetic band related to {\em
  all} the atoms in the unit cell we call magnetic ``super'' band. It
is highlighted in red in Fig.~\ref{fig:bandstr114} and consists of
band 1 of Mn, band 2 of As and band 3 of Ba in Tab.~\ref{tab:wf114}
and, hence, is defined by the symmetry labels
\begin{equation}
  \label{eq:5}
  \begin{array}{lll}
  \Gamma_1, \Gamma_2, \Gamma_3, \Gamma_4, & 2\Gamma_5, & \Gamma_3,
  \Gamma_4,\\
  M_1, M_2, M_3, M_4, & M_1, M_2, M_3, M_4, & M_5,\\
  2Z_5, & Z_1, Z_2, Z_3, Z_4, & Z_5,\\
  2A_5, &  2A_5, & A_1, A_2, \\   
  2R_1, &  2R_1, & R_1, \\
  2X_1, &  2X_1, & X_1 .\\
  \end{array}
\end{equation}

While the magnetic super band in BaMn$_2$As$_2$ satisfies the
condition (38) of Theorem 5 of Ref.~\cite{theoriewf} in all the point
of symmetry, there are little complications on the lines $\Sigma$ and
$\Delta$ in two branches: The four circles in
Fig.~\ref{fig:bandstr114} mark regions with unavoidable transitions
from $\Delta_2$ to $\Delta_1$, $\Delta_1$ to $\Delta_2$, $\Sigma_1$ to
$\Sigma_2$, and $\Sigma_2$ to $\Sigma_1$ symmetry. These transitions
clearly destroy the $P\overline{4}2_1c$ symmetry of the Wannier
functions. However, since these transitions only occur in two lines of
two branches fare away from the Fermi level, we assume that the
magnetic structure of BaMn$_2$As$_2$ may be described with high
accuracy in the space group $P\overline{4}2_1c$. Nevertheless, these
transitions produce an additional small distortion of the crystal
going beyond the displacements of the Mn atoms depicted in
Fig.~\ref{fig:structures} (b). This additional small distortion is not
considered in this paper. However, we should keep in mind (see
Note (v) of Table~\ref{tab:rep81}) 
that the
Wannier functions are {\em exactly} adapted to the space group
$P\overline{4} = \Gamma_qS^1_4$ (81) since in this space group the
mentioned complications on the lines $\Sigma$ and $\Delta$
disappear. The related exact magnetic group is a subgroup of
$M_{114}$~\gl{eq:4}. First, there exist two subgroups of $M_{114}$
defined by the two anti-unitary operations $\{KC_{2a}|000\}$ and
$\{KI|\frac{1}{2}\frac{1}{2}\frac{1}{2}\}$, respectively. A detailed
examination shows that only the group
\begin{equation}
  \label{eq:6}
  M_{81} = P\overline{4} + \{KC_{2a}|000\}P\overline{4}
\end{equation}
allows an additional distortion of the crystal.

\subsection{Time-inversion symmetry}
\label{sec:time_inversion}
The time-inversion symmetry is no essential object in
antiferromagnetic BaMn$_2$As$_2$: all the three space groups
$I4/mmm$~(139), $P\overline{4}2_1c$~(114), and $P\overline{4}$~(81)
possess one-dimensional representations allowing a {\em stable}
magnetic state with the magnetic group $M_{121}$~\gl{eq:2},
$M_{114}$~\gl{eq:4}, and $M_{81}$~\gl{eq:6}, respectively, see
Tables~\ref{tab:rep121},~\ref{tab:rep114}, and~\ref{tab:rep81} and the
notes to these tables. Consequently, time-inversion symmetry
influences neither the antiferromagnetic structure nor the structural
distortions in BaMn$_2$As$_2$ (as it is the case, for instance, in
BaFe$_2$As$_2$~\cite{bafe2as2}). Time-inversion symmetry only forbids
magnetic moments located at the As atoms, see
Note~(x) of Table~\ref{tab:wf114}.

\section{Physical Interpretation}
\label{sec:interpretation}
The existence and the properties of the roughly half-filled magnetic
super band in the band structure of BaMn$_2$As$_2$ yield to an
understanding of three phenomena that shall be considered in this section:
\begin{itemize}  
\item the experimentally observed~\cite{singh} antiferromagnetic order together
  with a structural distortion not yet experimentally found;
\item the semiconducting ground state; and
\item the different magnetic structures in BaMn$_2$As$_2$ and BaFe$_2$As$_2$.  
\end{itemize}

\subsection{The antiferromagnetic order and the 
structural distortion in \text{BaMn$_2$As$_2$}}
\label{sec:distortion}
In a material possessing a narrow, roughly half-filled magnetic band
or super band related to a magnetic group $M$, the NHM defines a
nonadiabatic Hamiltonian $H^n$ representing atomic-like electrons
(Section~\ref{sec:nhm}) within this band~\cite{enhm}.  An important
feature of $H^n$ is that it only commutes with the symmetry operations
of $M$, but {\em does not commute} with the remaining symmetry
operations of the paramagnetic group of the crystal (this follows from
the fact that the optimally-localized Wannier functions in a
magnetic band may be chosen symmetry-adapted to $M$, but cannot be
chosen symmetry-adapted to the complete paramagnetic group,
cf. Section 1 of Ref.~\cite{theoriewf}).

Thus, the electrons in such a narrow, roughly half-filled magnetic
band or super band may gain the nonadiabatic condensation energy
$\Delta E$ (Section~\ref{sec:nhm}) by condensing into an atomic-like
state only if the electrons really move in a potential with the
magnetic group $M$, that is, only if a magnetic structure with the
magnetic group $M$ really exists.  As a consequence, the electrons
{\em activate} in the nonadiabatic system a spin dependent exchange
mechanism producing a magnetic structure with the magnetic group
$M$~\cite{ea,ef}.

In the case of BaMn$_2$As$_2$, the group $I\overline{4}2m$ (121) is
the space group of the antiferromagnetic structure in {\em
  undistorted} BaMn$_2$As$_2$. However, within this group there does
not exist a magnetic band, see Sec.~\ref{sec:nonexistence}. Indeed, a
magnetic band, even a magnetic super band, exists in the space group
$P\overline{4}2_1c$ (114) of distorted BaMn$_2$As$_2$, see
Sec.~\ref{sec:existence}.

Hence, in BaMn$_2$As$_2$ the electron system cannot condense into the
atomic-like state by the production of the magnetic structure {\em
  alone} but must additionally produce a spatial distortion of the
crystal realizing - together with the magnetic structure - the
magnetic group $M_{114}$~\gl{eq:4}.  This is achieved by the
displacements of the Mn atoms depicted in Fig.~\ref{fig:structures}
(b). Consequently, the magnetically ordered ground state of
BaMn$_2$As$_2$ is accompanied by the displacements of the Mn
atoms depicted in Fig.~\ref{fig:structures} (b).

\subsection{The semiconducting ground state of BaMn$_2$As$_2$}
\label{sec:insulating}
The magnetic super band defines not only Wannier functions situated at
{\em all the} atoms of BaMn$_2$As$_2$, but it also comprises {\em all
  the Bloch states} at the Fermi level, see
Fig.~\ref{fig:bandstr114}. Thus, if the magnetic super band is {\em
  exactly} half filled, the nonadiabatic Hamiltonian $H^n$ produces
very specific atomic-like electrons: at {\em any} atom of
BaMn$_2$As$_2$ there exists a localized Wannier state occupied by
exactly one electron and besides these atomic-like electrons there
do not exist band-like electrons which would be able to transport
electrical current. Thus, $H^n$ possesses a semiconducting ground
state since the atomic-like state is separated from any band-like
state by the nonadiabatic condensation energy $\Delta E$ mentioned in
Section~\ref{sec:nhm}.  The experimental observation of an insulating
ground state in BaMn$_2$As$_2$ suggests that, indeed, the magnetic
super band is exactly half-filled.

\subsection{Different magnetic structures in \text{BaMn$_2$As$_2$} and
  \text{BaFe$_2$As$_2$}}
\label{sec:magneticstructure}
While both compounds BaMn$_2$As$_2$ and BaFe$_2$As$_2$ exhibit an
antiferromagnetic ordering below the respective N{\'e}el temperature,
the space groups of the magnetic structures are quite different: in
BaMn$_2$As$_2$ the space group of the magnetic structure is the
tetragonal group $P\overline{4}2_1c$ (114) with magnetic moments
oriented along the tetragonal $c$ axis, and in BaFe$_2$As$_2$ it is
the orthorhombic group $Cmca$ with magnetic moments orientated
perpendicular to the $c$ axis, see Fig.~1 of~\cite{singh} and Fig.~3
of~\cite{huang2}, respectively.

This surprising experimental observation can be understood comparing
the band structures of both compounds as given in
Fig.~\ref{fig:bandstr139} and Fig.~2 of Ref.~\cite{bafe2as2}. The band
structure of BaFe$_2$As$_2$ is very similar to the band structure of
BaMn$_2$As$_2$, the essential difference is the position of the Fermi
level: we may approximate the band structure of BaFe$_2$As$_2$ by the
band structure of BaMn$_2$As$_2$ by shifting the Fermi level upwards
by about 0.3 eV as it is indicated in Fig.~\ref{fig:bandstr139}.

A magnetic (super) band may be physically active only if the band is
nearly half-filled.  The band width of the (red) magnetic super band
in Fig.~\ref{fig:bandstr114} may be approximated by $2\sigma$, where
\begin{equation}
  \label{eq:7}
  \sigma = \sqrt{\frac{1}{N}\sum_{\bm k}(E_{\bm k} - E_F)^2}\ 
  \approx\  0.5 \text{eV}, 
\end{equation}
denotes the standard deviation of the $N = 6\cdot 10$ energy values
$E_{\bm k}$ in the six points of symmetry of the magnetic super
band. 

Thus, the Fermi level is shifted in BaFe$_2$As$_2$ nearly to the top
of the magnetic super band. Hence, in BaFe$_2$As$_2$ this band is far
from being half-filled and determines neither the magnetic structure
nor produces an isolating ground state in BaFe$_2$As$_2$. Instead, the
magnetic structure in BaFe$_2$As$_2$ is determined by the nearly
half-filled magnetic band presented in Fig.~3 of Ref.~\cite{bafe2as2}
which is related to the space group $Cmca$ of the magnetic structure
experimentally found in BaFe$_2$As$_2$~\cite{huang2}.

\section{Conclusions}
\label{sec:conclusions}
This paper emphasizes the importance of the nonadiabatic condensation
energy $\Delta E$ defined in Equation (2.20) of Ref.~\cite{enhm} (and
already mentioned in Section~\ref{sec:nhm}) which is evidently
responsible for the striking electronic features of
BaMn$_2$As$_2$. $\Delta E$ is released at the transition from an
adiabatic band-like motion of the electrons to the nonadiabatic
strongly-correlated atomic-like motion.

This finding is in accordance with former observations on a great
number of superconducting and magnetic materials (see Section 1
of~\cite{theoriewf}) suggesting that superconductivity and magnetism
are always connected with superconducting (Definition 22 of
Ref.~\cite{theoriewf}) and magnetic bands, respectively. Thus, in
superconducting and magnetic bands, the nonadiabatic condensation
energy $\Delta E$ may evidently produce superconductivity and
magnetism, respectively, and in some cases even a small band gap
semiconductor.

\acknowledgements
I am very indebted to Guido Schmitz for his support of my work. 

\appendix*
\onecolumngrid

\section{Group-theoretical tables}
This appendix provides Tables~\ref{tab:rep121} --~\ref{tab:rep81}
along with notes to the tables.

\label{sec:tables}

\begin{table}[b]
\caption{
  Character tables of the irreducible representations of the
  tetragonal space group $I\overline{4}2m = \Gamma^v_qD^{11}_{2d}$ (121) of the
  experimentally observed~\cite{singh} antiferromagnetic
  structure in BaMn$_2$As$_2$.  
  \label{tab:rep121}}
\begin{tabular}[t]{cccccccc}
\multicolumn{8}{c}{$\Gamma (000), Z (\frac{1}{2}\frac{1}{2}\overline{\frac{1}{2}})$}\\
 &  &  & $$ & $$ & $S^-_{4z}$ & $C_{2y}$ & $\sigma_{da}$\\
 & $K$ & $KI$ & $E$ & $C_{2z}$ & $S^+_{4z}$ & $C_{2x}$ & $\sigma_{db}$\\
\hline
$\Gamma_1, Z_1$ & (a) & (a) & 1 & 1 & 1 & 1 & 1\\
$\Gamma_2, Z_2$ & (a) & (a) & 1 & 1 & 1 & -1 & -1\\
$\Gamma_3, Z_3$ & (a) & (a) & 1 & 1 & -1 & 1 & -1\\
$\Gamma_4, Z_4$ & (a) & (a) & 1 & 1 & -1 & -1 & 1\\
$\Gamma_5, Z_5$ & (a) & (a) & 2 & -2 & 0 & 0 & 0\\
\hline\\
\end{tabular}\hspace{1cm}
\begin{tabular}[t]{cccccccc}
\multicolumn{8}{c}{$P (\frac{1}{4}\frac{1}{4}\frac{1}{4})$}\\
 &  &  & $$ & $$ & $S^-_{4z}$ & $C_{2y}$ & $\sigma_{da}$\\
 & $K$ & $KI$ & $E$ & $C_{2z}$ & $S^+_{4z}$ & $C_{2x}$ & $\sigma_{db}$\\
\hline
$P_1$ & (x) & (a) & 1 & 1 & 1 & 1 & 1\\
$P_2$ & (x) & (a) & 1 & 1 & 1 & -1 & -1\\
$P_3$ & (x) & (a) & 1 & 1 & -1 & 1 & -1\\
$P_4$ & (x) & (a) & 1 & 1 & -1 & -1 & 1\\
$P_5$ & (x) & (a) & 2 & -2 & 0 & 0 & 0\\
\hline\\
\end{tabular}\hspace{1cm}
\begin{tabular}[t]{ccccc}
\multicolumn{5}{c}{$X (00\frac{1}{2})$}\\
 & $E$ & $C_{2z}$ & $\sigma_{db}$ & $\sigma_{da}$\\
\hline
$X_1$ & 1 & 1 & 1 & 1\\
$X_3$ & 1 & 1 & -1 & -1\\
$X_2$ & 1 & -1 & 1 & -1\\
$X_4$ & 1 & -1 & -1 & 1\\
\hline\\
\end{tabular}\hspace{1cm}
\begin{tabular}[t]{ccc}
\multicolumn{3}{c}{$N (0\frac{1}{2}0)$}\\
 & $E$ & $C_{2y}$\\
\hline
$N_1$ & 1 & 1\\
$N_2$ & 1 & -1\\
\hline\\
\end{tabular}\hspace{1cm}

\hspace{1cm}
\ \\
\begin{flushleft}
Notes to Table~\ref{tab:rep121}
\end{flushleft}
\begin{enumerate}
\item The notations of the points of symmetry follow Fig. 3.10 (b) 
of Ref.~\cite{bc}.
\item The character tables are determined from Table 5.7 of
  Ref.~\protect\cite{bc}.  
\item $K$ denotes the operator of time inversion. The entry (a)
  indicates that the related corepresentations of the magnetic groups
  $I\overline{4}2m + \{K|000\}I\overline{4}2m$ and $I\overline{4}2m +
  \{KI|000\}I\overline{4}2m$ follow case (a) as defined in equation\ (7.3.45)
  of Ref.~\cite{bc} (and determined by equation\ (7.3.51) of
  Ref.~\cite{bc}). This information is interesting only in symmetry
  points invariant under the complete space group. 
  (x) indicates that $K$ does not leave invariant
  the point $P$. 
\item The one-dimensional
  representations at point $P$ would be possible representations
  of a {\em stable} antiferromagnetic state because they comply with
  the demands in Section III C of Ref.\ \cite{ea}.
\end{enumerate}
\end{table}


\begin{table}[!]
\caption{
Compatibility relations between the Brillouin zone for the space group
$I4/mmm$ (139) of paramagnetic BaMn$_2$As$_2$ and the Brillouin zone for the space group $I\overline{4}2m$ (121) of the
  antiferromagnetic structure in undistorted BaMn$_2$As$_2$.
\label{tab:falten139_121}
}
\begin{flushleft}
\begin{tabular}[t]{cccccccccc}
\multicolumn{10}{c}{$\Gamma (000)$}\\
\hline
$\Gamma^+_1$ & $\Gamma^+_2$ & $\Gamma^+_3$ & $\Gamma^+_4$ & $\Gamma^+_5$ & $\Gamma^-_1$ & $\Gamma^-_2$ & $\Gamma^-_3$ & $\Gamma^-_4$ & $\Gamma^-_5$\\
$\Gamma_1$ & $\Gamma_2$ & $\Gamma_3$ & $\Gamma_4$ & $\Gamma_5$ & $\Gamma_3$ & $\Gamma_4$ & $\Gamma_1$ & $\Gamma_2$ & $\Gamma_5$\\
\hline\\
\end{tabular}\hspace{1cm}
\begin{tabular}[t]{cccc}
\multicolumn{4}{c}{$N (0\frac{1}{2}0)$}\\
\hline
$N^+_1$ & $N^-_1$ & $N^+_2$ & $N^-_2$\\
$N_1$ & $N_1$ & $N_2$ & $N_2$\\
\hline\\
\end{tabular}\hspace{1cm}
\begin{tabular}[t]{cccccccc}
\multicolumn{8}{c}{$X (00\frac{1}{2})$}\\
\hline
$X^+_1$ & $X^+_2$ & $X^+_3$ & $X^+_4$ & $X^-_1$ & $X^-_2$ & $X^-_3$ & $X^-_4$\\
$X_1$ & $X_4$ & $X_3$ & $X_2$ & $X_3$ & $X_2$ & $X_1$ & $X_4$\\
\hline\\
\end{tabular}\hspace{1cm}
\begin{tabular}[t]{cccccccccc}
\multicolumn{10}{c}{$Z (\frac{1}{2}\frac{1}{2}\overline{\frac{1}{2}})$}\\
\hline
$Z^+_1$ & $Z^+_2$ & $Z^+_3$ & $Z^+_4$ & $Z^+_5$ & $Z^-_1$ & $Z^-_2$ & $Z^-_3$ & $Z^-_4$ & $Z^-_5$\\
$Z_1$ & $Z_2$ & $Z_3$ & $Z_4$ & $Z_5$ & $Z_3$ & $Z_4$ & $Z_1$ & $Z_2$ & $Z_5$\\
\hline\\
\end{tabular}\hspace{1cm}
\begin{tabular}[t]{ccccc}
\multicolumn{5}{c}{$P (\frac{1}{4}\frac{1}{4}\frac{1}{4})$}\\
\hline
$P_1$ & $P_2$ & $P_3$ & $P_4$ & $P_5$\\
$P_1$ & $P_2$ & $P_3$ & $P_4$ & $P_5$\\
\hline\\
\end{tabular}\hspace{1cm}
\end{flushleft}
\ \\
\begin{flushleft}
Notes to Table~\ref{tab:falten139_121}
\end{flushleft}
\begin{enumerate}
\item The Brillouin zone for $I\overline{4}2m$ is identical to the Brillouin zone for $I4/mmm$.
\item The upper rows list the representations of the little groups of the
  points of symmetry in the Brillouin zone for $I4/mmm$.  The lower rows list
  representations of these groups in $I\overline{4}2m$. 
  
  The representations in the same column are compatible in the
  following sense: Bloch functions that are basis functions of a
  representation $\bm{D}_i$ in the upper row can be unitarily transformed into
  the basis functions of the representation given below $\bm{D}_i$.
\item The notations of the representations are defined in
 Table 2 of Ref.~\cite{bafe2as2}  and Table~\ref{tab:rep121}, respectively.
\end{enumerate}
\end{table}


\begin{table}
\caption{
Representations at the points of symmetry in the space group $I\overline{4}2m =
\Gamma^v_qD^{11}_{2d}$ (121) of all the energy bands of antiferromagnetic 
BaMn$_2$As$_2$ with symmetry-adapted and optimally  
localized Wannier functions centered at the Mn, As, and Ba atoms, respectively. 
\label{tab:wf121}}
\begin{tabular}{ccccccccc}
\bf{Mn} & Mn($\frac{1}{4}\frac{3}{4}\frac{1}{2}$) & Mn($\frac{3}{4}\frac{1}{4}\frac{1}{2}$) & $KI$ & $\Gamma$ & $P$ & $Z$ & $X$ & $N$\\
\hline
Band 1 & $\bm{d}_{1}$ & $\bm{d}_{1}$ & OK & $\Gamma_1$ + $\Gamma_2$ & $P_5$ & $Z_3$ + $Z_4$ & $X_2$ + $X_4$ & $N_1$ + $N_2$\\
Band 2 & $\bm{d}_{2}$ & $\bm{d}_{4}$ & OK & $\Gamma_5$ & $P_3$ + $P_4$ & $Z_5$ & $X_1$ + $X_3$ & $N_1$ + $N_2$\\
Band 3 & $\bm{d}_{3}$ & $\bm{d}_{3}$ & OK & $\Gamma_3$ + $\Gamma_4$ & $P_5$ & $Z_1$ + $Z_2$ & $X_2$ + $X_4$ & $N_1$ + $N_2$\\
Band 4 & $\bm{d}_{4}$ & $\bm{d}_{2}$ & OK & $\Gamma_5$ & $P_1$ + $P_2$ & $Z_5$ & $X_1$ + $X_3$ & $N_1$ + $N_2$\\
\hline\\
\end{tabular}\hspace{1cm}
\begin{tabular}{ccccccccc}
\bf{As} & As($zz0$) & As($\overline{z}\overline{z}0$) & $KI$ & $\Gamma$ & $P$ & $Z$ & $X$ & $N$\\
\hline
Band 1 & $\bm{d}_{1}$ & $\bm{d}_{1}$ & OK & $\Gamma_1$ + $\Gamma_4$ & $P_1$ + $P_4$ & $Z_1$ + $Z_4$ & 2$X_1$ & $N_1$ + $N_2$\\
Band 2 & $\bm{d}_{2}$ & $\bm{d}_{2}$ & OK & $\Gamma_2$ + $\Gamma_3$ & $P_2$ + $P_3$ & $Z_2$ + $Z_3$ & 2$X_3$ & $N_1$ + $N_2$\\
Band 3 & $\bm{d}_{3}$ & $\bm{d}_{4}$ & $-$ & $\Gamma_5$ & $P_5$ & $Z_5$ & $X_2$ + $X_4$ & $N_1$ + $N_2$\\
\hline\\
\end{tabular}\hspace{1cm}
\begin{tabular}{cccccccc}
\bf{Ba} & Ba($000$) & $KI$ & $\Gamma$ & $P$ & $Z$ & $X$ & $N$\\
\hline
Band 1 & $\bm{d}_{1}$ & OK & $\Gamma_1$ & $P_1$ & $Z_1$ & $X_1$ & $N_1$\\
Band 2 & $\bm{d}_{2}$ & OK & $\Gamma_2$ & $P_2$ & $Z_2$ & $X_3$ & $N_2$\\
Band 3 & $\bm{d}_{3}$ & OK & $\Gamma_3$ & $P_3$ & $Z_3$ & $X_3$ & $N_1$\\
Band 4 & $\bm{d}_{4}$ & OK & $\Gamma_4$ & $P_4$ & $Z_4$ & $X_1$ & $N_2$\\
\hline\\
\end{tabular}\hspace{1cm}
\ \\
\begin{flushleft}
Notes to Table~\ref{tab:wf121}
\end{flushleft}
\begin{enumerate}
\item $z = 0.36\ldots$\ \cite{singh}; the exact value of $z$ is
  meaningless in this table. 
\item The antiferromagnetic structure of undistorted BaMn$_2$As$_2$ has the space group $I\overline{4}2m$ and the magnetic
group $M = I\overline{4}2m + \{KI|000\}I\overline{4}2m$ with
$K$ denoting the operator of time-inversion.
\item Each row defines a band with Bloch functions that can be
  unitarily transformed into Wannier functions being
\begin{itemize}
\item as well localized as possible; 
\item centered at the stated atoms;
\item and symmetry-adapted to the space group $I\overline{4}2m$ of the
  antiferromagnetic structure in undistorted BaMn$_2$As$_2$.
\end{itemize}
\item The notations of the representations are defined in Table~\ref{tab:rep121}.
\item The bands are determined following Theorem 5 of Ref.\
  \cite{theoriewf}.
\item The Wannier functions at the Mn, As or Ba atom listed in the
  upper row belong to the representation $\bm{d}_i$  included below the atom. 
\item The $\bm{d}_i$ denote the one-dimensional representations of the
  ``point groups of the positions'' of the Mn, As and Ba atom (Definition
  12 of Ref.\ \cite{theoriewf}), $S_4$, $C_{2v}$, and $D_{2d}$, respectively,
  as defined by the tables
\begin{center}
\begin{tabular}[t]{ccccc}
\multicolumn{5}{c}{Mn atoms}\\
\\
 & $E$ & $S^+_{4z}$ & $C_{2z}$ & $S^-_{4z}$\\
\hline
$\bm{d}_{1}$ & 1 & 1 & 1 & 1\\
$\bm{d}_{2}$ & 1 & i & -1 & -i\\
$\bm{d}_{3}$ & 1 & -1 & 1 & -1\\
$\bm{d}_{4}$ & 1 & -i & -1 & i\\
\hline\\
\end{tabular}\hspace{1cm}
\begin{tabular}[t]{ccccc}
\multicolumn{5}{c}{As atoms}\\
\\
 & $E$ & $C_{2z}$ & $\sigma_{da}$ & $\sigma_{db}$\\
\hline
$\bm{d}_{1}$ & 1 & 1 & 1 & 1\\
$\bm{d}_{2}$ & 1 & 1 & -1 & -1\\
$\bm{d}_{3}$ & 1 & -1 & 1 & -1\\
$\bm{d}_{4}$ & 1 & -1 & -1 & 1\\
\hline\\
\end{tabular}\hspace{1cm}
\begin{tabular}[t]{cccccc}
\multicolumn{6}{c}{Ba atom}\\
 & $$ & $$ & $S^-_{4z}$ & $C_{2y}$ & $\sigma_{da}$\\
 & $E$ & $C_{2z}$ & $S^+_{4z}$ & $C_{2x}$ & $\sigma_{db}$\\
\hline
$\bm{d}_{1}$ & 1 & 1 & 1 & 1 & 1\\
$\bm{d}_{2}$ & 1 & 1 & 1 & -1 & -1\\
$\bm{d}_{3}$ & 1 & 1 & -1 & 1 & -1\\
$\bm{d}_{4}$ & 1 & 1 & -1 & -1 & 1\\
\hline\\
\end{tabular}\hspace{1cm}
\end{center}
\item The entry ``OK'' indicates whether the Wannier functions may even be
  chosen symmetry-adapted to the magnetic group $M = I\overline{4}2m +
  \{KI|000\}I\overline{4}2m$ of undistorted BaMn$_2$As$_2$, see
  Theorem 7 of Ref.\ \cite{theoriewf}.
\item Hence, all the listed bands except for band 3 of As form
  magnetic bands as defined by Definition 16 of Ref.\
  \cite{theoriewf}.
\item Each band consists of one or two branches (Definition 2 of Ref.\
  \cite{theoriewf}) depending on the number of the related atoms in
  the unit cell.   
\end{enumerate}
\end{table}


\begin{table}[t]
\caption{
  Character tables of the irreducible representations of the
  tetragonal space group $P\overline{4}2_1c = \Gamma_qD^{4}_{2d}$ (114) of the
  experimentally observed~\cite{singh} antiferromagnetic
  structure in distorted BaMn$_2$As$_2$.
  \label{tab:rep114}}
\begin{tabular}[t]{cccccccc}
\multicolumn{8}{c}{$\Gamma (000)$}\\
 &  &  & $$ & $$ & $\{S^-_{4z}|000\}$ & $\{C_{2y}|\frac{1}{2}\frac{1}{2}\frac{1}{2}\}$ & $\{\sigma_{da}|\frac{1}{2}\frac{1}{2}\frac{1}{2}\}$\\
 & $K$ & $\{KI|\frac{1}{2}\frac{1}{2}\frac{1}{2}\}$ & $\{E|000\}$ & $\{C_{2z}|000\}$ & $\{S^+_{4z}|000\}$ & $\{C_{2x}|\frac{1}{2}\frac{1}{2}\frac{1}{2}\}$ & $\{\sigma_{db}|\frac{1}{2}\frac{1}{2}\frac{1}{2}\}$\\
\hline
$\Gamma_1$ & (a) & (a) & 1 & 1 & 1 & 1 & 1\\
$\Gamma_2$ & (a) & (a) & 1 & 1 & 1 & -1 & -1\\
$\Gamma_3$ & (a) & (a) & 1 & 1 & -1 & 1 & -1\\
$\Gamma_4$ & (a) & (a) & 1 & 1 & -1 & -1 & 1\\
$\Gamma_5$ & (a) & (a) & 2 & -2 & 0 & 0 & 0\\
\hline\\
\end{tabular}\hspace{1cm}
\begin{tabular}[t]{ccccccccccccc}
\multicolumn{13}{c}{$M (\frac{1}{2}\frac{1}{2}0)$}\\
 &  &  & $$ & $$ & $$ & $$ & $\{\sigma_{da}|\frac{1}{2}\frac{3}{2}\frac{1}{2}\}$ & $\{\sigma_{db}|\frac{1}{2}\frac{3}{2}\frac{1}{2}\}$ & $\{S^-_{4z}|010\}$ & $\{S^+_{4z}|010\}$ & $\{C_{2x}|\frac{1}{2}\frac{1}{2}\frac{1}{2}\}$ & $\{C_{2y}|\frac{1}{2}\frac{1}{2}\frac{1}{2}\}$\\
 & $K$ & $\{KI|\frac{1}{2}\frac{1}{2}\frac{1}{2}\}$ & $\{E|000\}$ & $\{C_{2z}|010\}$ & $\{C_{2z}|000\}$ & $\{E|010\}$ & $\{\sigma_{db}|\frac{1}{2}\frac{1}{2}\frac{1}{2}\}$ & $\{\sigma_{da}|\frac{1}{2}\frac{1}{2}\frac{1}{2}\}$ & $\{S^+_{4z}|000\}$ & $\{S^-_{4z}|000\}$ & $\{C_{2y}|\frac{1}{2}\frac{3}{2}\frac{1}{2}\}$ & $\{C_{2x}|\frac{1}{2}\frac{3}{2}\frac{1}{2}\}$\\
\hline
$M_1$ & (c) & (a) & 1 & 1 & -1 & -1 & 1 & -1 & i & -i & i & -i\\
$M_2$ & (c) & (a) & 1 & 1 & -1 & -1 & 1 & -1 & -i & i & -i & i\\
$M_3$ & (c) & (a) & 1 & 1 & -1 & -1 & -1 & 1 & -i & i & i & -i\\
$M_4$ & (c) & (a) & 1 & 1 & -1 & -1 & -1 & 1 & i & -i & -i & i\\
$M_5$ & (a) & (a) & 2 & -2 & 2 & -2 & 0 & 0 & 0 & 0 & 0 & 0\\
\hline\\
\end{tabular}\hspace{1cm}
\begin{tabular}[t]{ccccccccccccc}
\multicolumn{13}{c}{$Z (00\frac{1}{2})$}\\
 &  &  & $$ & $$ & $$ & $$ & $\{C_{2y}|\frac{1}{2}\frac{1}{2}\frac{3}{2}\}$ & $\{C_{2x}|\frac{1}{2}\frac{1}{2}\frac{3}{2}\}$ & $\{S^-_{4z}|001\}$ & $\{S^+_{4z}|001\}$ & $\{\sigma_{da}|\frac{1}{2}\frac{1}{2}\frac{1}{2}\}$ & $\{\sigma_{db}|\frac{1}{2}\frac{1}{2}\frac{1}{2}\}$\\
 & $K$ & $\{KI|\frac{1}{2}\frac{1}{2}\frac{1}{2}\}$ & $\{E|000\}$ & $\{C_{2z}|001\}$ & $\{C_{2z}|000\}$ & $\{E|001\}$ & $\{C_{2x}|\frac{1}{2}\frac{1}{2}\frac{1}{2}\}$ & $\{C_{2y}|\frac{1}{2}\frac{1}{2}\frac{1}{2}\}$ & $\{S^+_{4z}|000\}$ & $\{S^-_{4z}|000\}$ & $\{\sigma_{db}|\frac{1}{2}\frac{1}{2}\frac{3}{2}\}$ & $\{\sigma_{da}|\frac{1}{2}\frac{1}{2}\frac{3}{2}\}$\\
\hline
$Z_1$ & (c) & (a) & 1 & 1 & -1 & -1 & 1 & -1 & i & -i & i & -i\\
$Z_2$ & (c) & (a) & 1 & 1 & -1 & -1 & 1 & -1 & -i & i & -i & i\\
$Z_3$ & (c) & (a) & 1 & 1 & -1 & -1 & -1 & 1 & -i & i & i & -i\\
$Z_4$ & (c) & (a) & 1 & 1 & -1 & -1 & -1 & 1 & i & -i & -i & i\\
$Z_5$ & (a) & (a) & 2 & -2 & 2 & -2 & 0 & 0 & 0 & 0 & 0 & 0\\
\hline\\
\end{tabular}\hspace{1cm}
\begin{tabular}[t]{ccccccccccccc}
\multicolumn{13}{c}{$A (\frac{1}{2}\frac{1}{2}\frac{1}{2})$}\\
 &  &  & $$ & $$ & $$ & $$ & $\{S^-_{4z}|000\}$ & $\{S^-_{4z}|001\}$ & $\{\sigma_{da}|\frac{1}{2}\frac{1}{2}\frac{1}{2}\}$ & $\{C_{2x}|\frac{1}{2}\frac{1}{2}\frac{1}{2}\}$ & $\{\sigma_{da}|\frac{1}{2}\frac{1}{2}\frac{3}{2}\}$ & $\{C_{2x}|\frac{1}{2}\frac{1}{2}\frac{3}{2}\}$\\
 & $K$ & $\{KI|\frac{1}{2}\frac{1}{2}\frac{1}{2}\}$ & $\{E|000\}$ & $\{C_{2z}|000\}$ & $\{E|001\}$ & $\{C_{2z}|001\}$ & $\{S^+_{4z}|000\}$ & $\{S^+_{4z}|001\}$ & $\{\sigma_{db}|\frac{1}{2}\frac{1}{2}\frac{1}{2}\}$ & $\{C_{2y}|\frac{1}{2}\frac{1}{2}\frac{1}{2}\}$ & $\{\sigma_{db}|\frac{1}{2}\frac{1}{2}\frac{3}{2}\}$ & $\{C_{2y}|\frac{1}{2}\frac{1}{2}\frac{3}{2}\}$\\
\hline
$A_1$ & (c) & (a) & 1 & 1 & -1 & -1 & -1 & 1 & i & -i & -i & i\\
$A_2$ & (c) & (a) & 1 & 1 & -1 & -1 & -1 & 1 & -i & i & i & -i\\
$A_3$ & (c) & (a) & 1 & 1 & -1 & -1 & 1 & -1 & -i & -i & i & i\\
$A_4$ & (c) & (a) & 1 & 1 & -1 & -1 & 1 & -1 & i & i & -i & -i\\
$A_5$ & (b) & (a) & 2 & -2 & -2 & 2 & 0 & 0 & 0 & 0 & 0 & 0\\
\hline\\
\end{tabular}\hspace{1cm}
\begin{tabular}[t]{cccccc}
\multicolumn{6}{c}{$R (0\frac{1}{2}\frac{1}{2})$}\\
 & $$ & $$ & $\{C_{2y}|\frac{1}{2}\frac{1}{2}\frac{3}{2}\}$ & $\{C_{2z}|001\}$ & $\{C_{2x}|\frac{1}{2}\frac{1}{2}\frac{3}{2}\}$\\
 & $\{E|000\}$ & $\{E|001\}$ & $\{C_{2y}|\frac{1}{2}\frac{1}{2}\frac{1}{2}\}$ & $\{C_{2z}|000\}$ & $\{C_{2x}|\frac{1}{2}\frac{1}{2}\frac{1}{2}\}$\\
\hline
$R_1$ & 2 & -2 & 0 & 0 & 0\\
\hline\\
\end{tabular}\hspace{1cm}
\begin{tabular}[t]{cccccc}
\multicolumn{6}{c}{$X (0\frac{1}{2}0)$}\\
 & $$ & $$ & $\{C_{2y}|\frac{1}{2}\frac{3}{2}\frac{1}{2}\}$ & $\{C_{2z}|010\}$ & $\{C_{2x}|\frac{1}{2}\frac{3}{2}\frac{1}{2}\}$\\
 & $\{E|000\}$ & $\{E|010\}$ & $\{C_{2y}|\frac{1}{2}\frac{1}{2}\frac{1}{2}\}$ & $\{C_{2z}|000\}$ & $\{C_{2x}|\frac{1}{2}\frac{1}{2}\frac{1}{2}\}$\\
\hline
$X_1$ & 2 & -2 & 0 & 0 & 0\\
\hline\\
\end{tabular}\hspace{1cm}
\ \\
\begin{flushleft}
Notes to Table~\ref{tab:rep114}
\end{flushleft}
\begin{enumerate}
\item The notations of the points of symmetry follow Fig. 3.9 
of Ref.~\cite{bc}.
\item The character tables are determined from Table 5.7 of
  Ref.~\protect\cite{bc}.  
\item $K$ denotes the operator of time inversion. The entries (a), (b)
  and (c) indicate whether the related co-representations of the
  magnetic groups $P\overline{4}2_1c + \{K|000\}P\overline{4}2_1c$ and
  $P\overline{4}2_1c +
  \{KI|\frac{1}{2}\frac{1}{2}\frac{1}{2}\}P\overline{4}2_1c$ follow
  case (a), (b) or (c) as defined in equations\ (7.3.45),
  (7.3.46) and (7.3.47), respectively, of Ref.~\cite{bc} (and
  determined by equation\ (7.3.51) of Ref.~\cite{bc}). This
  information is interesting only in symmetry points invariant under
  the complete space group.
\item The entries (a) and (c) for $K$ and
  $\{KI|\frac{1}{2}\frac{1}{2}\frac{1}{2}\}$ show that all the one-dimensional
  representations at $M$, $Z$, or $A$ are possible representations
  of a {\em stable} antiferromagnetic state, see Appendix A of Ref.\
  \cite{josla2cuo4} or Section III C of Ref.\ \cite{ea}. 
\end{enumerate}
\end{table}

\begin{table}[!]
\caption{
Compatibility relations between the Brillouin zone for the space group
$I4/mmm$ (139) of tetragonal
paramagnetic BaMn$_2$As$_2$ and the Brillouin zone for the space group $P\overline{4}2_1c$ (114) of the
  antiferromagnetic
  structure in distorted BaMn$_2$As$_2$.
\label{tab:falten139_114}
}
\begin{tabular}[t]{cccccccccc}
\multicolumn{10}{c}{$\Gamma (000)$}\\
\hline
$\Gamma^+_1$ & $\Gamma^+_2$ & $\Gamma^+_3$ & $\Gamma^+_4$ & $\Gamma^+_5$ & $\Gamma^-_1$ & $\Gamma^-_2$ & $\Gamma^-_3$ & $\Gamma^-_4$ & $\Gamma^-_5$\\
$\Gamma_1$ & $\Gamma_2$ & $\Gamma_3$ & $\Gamma_4$ & $\Gamma_5$ & $\Gamma_3$ & $\Gamma_4$ & $\Gamma_1$ & $\Gamma_2$ & $\Gamma_5$\\
\hline\\
\end{tabular}\hspace{1cm}
\begin{tabular}[t]{cccccccc}
\multicolumn{8}{c}{$X (00\frac{1}{2})$}\\
\hline
$X^+_1$ & $X^+_2$ & $X^+_3$ & $X^+_4$ & $X^-_1$ & $X^-_2$ & $X^-_3$ & $X^-_4$\\
$M_5$ & $M_1$ + $M_2$ & $M_5$ & $M_3$ + $M_4$ & $M_5$ & $M_3$ + $M_4$ & $M_5$ & $M_1$ + $M_2$\\
\hline\\
\end{tabular}\hspace{1cm}
\begin{tabular}[t]{cccccccccc}
\multicolumn{10}{c}{$Z (\frac{1}{2}\frac{1}{2}\overline{\frac{1}{2}})$}\\
\hline
$Z^+_1$ & $Z^+_2$ & $Z^+_3$ & $Z^+_4$ & $Z^+_5$ & $Z^-_1$ & $Z^-_2$ & $Z^-_3$ & $Z^-_4$ & $Z^-_5$\\
$\Gamma_2$ & $\Gamma_1$ & $\Gamma_4$ & $\Gamma_3$ & $\Gamma_5$ & $\Gamma_4$ & $\Gamma_3$ & $\Gamma_2$ & $\Gamma_1$ & $\Gamma_5$\\
\hline\\
\end{tabular}\hspace{1cm}
\begin{tabular}[t]{ccccc}
\multicolumn{5}{c}{$P (\frac{1}{4}\frac{1}{4}\frac{1}{4})$}\\
\hline
$P_1$ & $P_2$ & $P_3$ & $P_4$ & $P_5$\\
$A_3$ + $A_4$ & $A_3$ + $A_4$ & $A_1$ + $A_2$ & $A_1$ + $A_2$ & 2$A_5$\\
\hline\\
\end{tabular}\hspace{1cm}
\begin{tabular}[t]{ccccc}
\multicolumn{5}{c}{${\Lambda_M} (\frac{1}{4}\frac{1}{4}\overline{\frac{1}{4}})$\quad line $\Lambda$}\\
\hline
$\Lambda_1$ & $\Lambda_2$ & $\Lambda_3$ & $\Lambda_4$ & $\Lambda_5$\\
$Z_5$ & $Z_5$ & $Z_5$ & $Z_5$ & $Z_1$ + $Z_2$ + $Z_3$ + $Z_4$\\
\hline\\
\end{tabular}\hspace{1cm}
\ \\
\begin{flushleft}
Notes to Table~\ref{tab:falten139_114}
\end{flushleft}
\begin{enumerate}
\item The Brillouin zone for $P\overline{4}2_1c$ lies within the Brillouin zone for $I4/mmm$.
\item The upper rows list the representations of the little groups of the
  points of symmetry in the Brillouin zone for $I4/mmm$ and the lower rows list
  representations of the little groups of the related points of symmetry in
  the Brillouin zone for $P\overline{4}2_1c$.
  
  The representations in the same column are compatible in the
  following sense: Bloch functions that are basis functions of a
  representation $\bm{D}_i$ in the upper row can be unitarily transformed into
  the basis functions of the representation given below $\bm{D}_i$.
\item The notations of the representations are defined in
  Table 2 of Ref.\ \cite{bafe2as2} and Table~\ref{tab:rep114}, respectively.
\item $\Lambda_M (\frac{1}{4}\frac{1}{4}\overline{\frac{1}{4}})$
  denotes the midpoint between $\Gamma$ and $Z$ in the Brillouin zone for $I4/mmm$.  
\item The representations on the line $\Lambda$ in the Brillouin zone
  for $I4/mmm$ are simple: the branch connecting $\Gamma_5$ and $Z_5$
  in Fig.\ref{fig:bandstr139} is labeled by the two-dimensional
  representation $\Lambda_5$, all the other branches are labeled by one
  of the one-dimensional representations $\Lambda_1$, $\Lambda_2$,
  $\Lambda_3$, or $\Lambda_4$.
\item The compatibility relations are determined in the way described in
  great detail in Ref.~\cite{eabf}.
\end{enumerate}
\end{table}



\begin{table}
\caption{
Representations at the points of symmetry in the space group $P\overline{4}2_1c$ (114) of all
the energy bands of {\em distorted} antiferromagnetic 
BaMn$_2$As$_2$ with symmetry-adapted and optimally  
localized Wannier functions centered at the Mn, As, or Ba atoms, respectively. 
\label{tab:wf114}}
\begin{tabular}[t]{cccccccc}
{\bf Mn} & Mn($\frac{1}{2}0\frac{1}{4}$) & Mn($0\frac{1}{2}\frac{1}{4}$) & Mn($1\frac{1}{2}\frac{3}{4}$) & Mn($\frac{1}{2}1\frac{3}{4}$) & $\{KI|\frac{1}{2}\frac{1}{2}\frac{1}{2}\}$ & $\Gamma$ & $M$\\
\hline
Band 1 & $\bm{d}_{1}$ & $\bm{d}_{1}$ & $\bm{d}_{1}$ & $\bm{d}_{1}$ & OK & $\Gamma_1$ + $\Gamma_2$ + $\Gamma_3$ + $\Gamma_4$ & $M_1$ + $M_2$ + $M_3$ + $M_4$\\
Band 2 & $\bm{d}_{2}$ & $\bm{d}_{2}$ & $\bm{d}_{2}$ & $\bm{d}_{2}$ & $*$ & 2$\Gamma_5$ & 2$M_5$\\
\hline\\
\end{tabular}\hspace{1cm}
\begin{tabular}[t]{ccccc}
\multicolumn{5}{c}{$(continued)$}\\
{\bf Mn} & $Z$ & $A$ & $R$ & $X$\\
\hline
Band 1 & 2$Z_5$ & 2$A_5$ & 2$R_1$ & 2$X_1$\\
Band 2 & $Z_1$ + $Z_2$ + $Z_3$ + $Z_4$ & $A_1$ + $A_2$ + $A_3$ + $A_4$ & 2$R_1$ & 2$X_1$\\
\hline\\
\\
\end{tabular}\hspace{1cm}
\begin{tabular}[t]{cccccccc}
{\bf As} &  As($00z$) & As($00\overline{z}$) &
As($\frac{1}{2}\frac{1}{2},\frac{1}{2}+z$) & 
As($\frac{1}{2}\frac{1}{2},\frac{1}{2}-z$) & $\{KI|\frac{1}{2}\frac{1}{2}\frac{1}{2}\}$ & $\Gamma$ & $M$\\
\hline
Band 1 & $\bm{d}_{1}$ & $\bm{d}_{1}$ & $\bm{d}_{1}$ & $\bm{d}_{1}$ & OK & $\Gamma_1$ + $\Gamma_2$ + $\Gamma_3$ + $\Gamma_4$ & 2$M_5$\\
Band 2 & $\bm{d}_{2}$ & $\bm{d}_{2}$ & $\bm{d}_{2}$ & $\bm{d}_{2}$ & $*$ & 2$\Gamma_5$ & $M_1$ + $M_2$ + $M_3$ + $M_4$\\
\hline\\
\end{tabular}\hspace{1cm}
\begin{tabular}[t]{ccccc}
\multicolumn{5}{c}{$(continued)$}\\
{\bf As} & $Z$ & $A$ & $R$ & $X$\\
\hline
Band 1 & 2$Z_5$ & $A_1$ + $A_2$ + $A_3$ + $A_4$ & 2$R_1$ & 2$X_1$\\
Band 2 & $Z_1$ + $Z_2$ + $Z_3$ + $Z_4$ & 2$A_5$ & 2$R_1$ & 2$X_1$\\
\hline\\
\\
\end{tabular}\hspace{1cm}
\begin{tabular}[t]{cccccccccc}
{\bf Ba} & Ba($000$) & Ba($\frac{1}{2}\frac{1}{2}\frac{1}{2}$) & $\{KI|\frac{1}{2}\frac{1}{2}\frac{1}{2}\}$ & $\Gamma$ & $M$ & $Z$ & $A$ & $R$ & $X$\\
\hline
Band 1 & $\bm{d}_{1}$ & $\bm{d}_{1}$ & OK & $\Gamma_1$ + $\Gamma_2$ & $M_5$ & $Z_5$ & $A_3$ + $A_4$ & $R_1$ & $X_1$\\
Band 2 & $\bm{d}_{2}$ & $\bm{d}_{4}$ & OK & $\Gamma_5$ & $M_1$ + $M_4$ & $Z_1$ + $Z_4$ & $A_5$ & $R_1$ & $X_1$\\
Band 3 & $\bm{d}_{3}$ & $\bm{d}_{3}$ & OK & $\Gamma_3$ + $\Gamma_4$ & $M_5$ & $Z_5$ & $A_1$ + $A_2$ & $R_1$ & $X_1$\\
Band 4 & $\bm{d}_{4}$ & $\bm{d}_{2}$ & OK & $\Gamma_5$ & $M_2$ + $M_3$ & $Z_2$ + $Z_3$ & $A_5$ & $R_1$ & $X_1$\\
\hline\\
\end{tabular}
\begin{flushleft}
Notes to Table~\ref{tab:wf114}
\end{flushleft}
\begin{enumerate}
\item $z = 0.36\ldots$\ \cite{singh}; the exact value of $z$ is
  meaningless in this table. 
\item The space group $P\overline{4}2_1c$ leaves invariant the
  experimentally observed \cite{singh} antiferromagnetic structure and
  defines the distortion of BaMn$_2$As$_2$ that possesses the magnetic
  super band consisting of band 1 of Mn, band 2 of As, and band 3 of Ba.
\item The appertaining magnetic group reads as $M = P\overline{4}2_1c +
  \{KI|\frac{1}{2}\frac{1}{2}\frac{1}{2}\}P\overline{4}2_1c$,
where $K$ still denotes the operator of time-inversion.
\item The notations of the representations are defined in Table~\ref{tab:rep114}.
\item The bands are determined following Theorem 5 of Ref.\
  \cite{theoriewf}.
\item Each row defines a band with Bloch functions that can be
  unitarily transformed into Wannier functions being
\begin{itemize}
\item as well localized as possible; 
\item centered at the stated atoms; and
\item symmetry-adapted to $P\overline{4}2_1c$.
\end{itemize}
\end{enumerate}
\end{table}
\begin{table}
\begin{flushleft}
Notes to Table~\ref{tab:wf114} (continued)
\end{flushleft}
\begin{enumerate}
\setcounter{enumi}{6}
\item The Wannier functions at the Mn, As or Ba atom listed in the
  upper row belong to the representation $\bm{d}_i$  included below the atom. 
\item The $\bm{d}_i$ denote the representations of the
  ``point groups of the positions'' of the Mn, As and Ba atoms (Definition
  12 of Ref.\ \cite{theoriewf}), $C_2$, $C_2$, and $S_4$, respectively,
  as defined by the tables
\begin{center}
\begin{tabular}[t]{ccc}
\multicolumn{3}{c}{Mn atoms}\\
 & $\{E|000\}$ & $\{C_{2z}|000\}$\\
\hline
$\bm{d}_{1}$ & 1 & 1\\
$\bm{d}_{2}$ & 1 & -1\\
\hline\\
\end{tabular}\hspace{1cm}
\begin{tabular}[t]{ccc}
\multicolumn{3}{c}{As atoms}\\
 & $\{E|000\}$ & $\{C_{2z}|000\}$\\
\hline
$\bm{d}_{1}$ & 1 & 1\\
$\bm{d}_{2}$ & 1 & -1\\
\hline\\
\end{tabular}\hspace{1cm}
\begin{tabular}[t]{ccccc}
\multicolumn{5}{c}{Ba atoms}\\
 & $\{E|000\}$ & $\{S^+_{4z}|000\}$ & $\{C_{2z}|000\}$ & $\{S^-_{4z}|000\}$\\
\hline
$\bm{d}_{1}$ & 1 & 1 & 1 & 1\\
$\bm{d}_{2}$ & 1 & i & -1 & -i\\
$\bm{d}_{3}$ & 1 & -1 & 1 & -1\\
$\bm{d}_{4}$ & 1 & -i & -1 & i\\
\hline\\
\end{tabular}\hspace{1cm}
\end{center}
\item The entry ``OK'' indicates whether the Wannier functions may
  even be chosen symmetry-adapted to the magnetic group $M =
  P\overline{4}2_1c +
  \{KI|\frac{1}{2}\frac{1}{2}\frac{1}{2}\}P\overline{4}2_1c$, see
  Theorem 7 of Ref.\ \cite{theoriewf}.
\item 
\label{item:asterisk}
The asterisk ``$*$'' indicates that the Wannier functions may be
  chosen symmetry-adapted to the magnetic group $M$, but they do not
  allow that the magnetic moments are situated at the appertaining atoms. This
  complication (which has not yet been considered in Ref.\
  \cite{theoriewf}) may (but does not necessarily) occur only if the
  representations of the space group at point $\Gamma$ are not
  one-dimensional as it is the case in band 2 of both Mn and As, and
  in bands 2 and 4 of Ba. Consider, for example, band 2 of Mn and the
  two Mn($\frac{1}{2}0\frac{1}{4}$) and Mn($\frac{1}{2}1\frac{3}{4}$)
  atoms. The magnetic moments at the two positions $A$ and $B$ of these atoms are
  anti-parallel. Thus, the two Wannier functions $w_A(\vec r)$ and
  $w_B(\vec r)$ at these positions are complex conjugate,
  $w_A(\vec r) = w^*_B(\vec r),$ and, hence, belong to
  co-representations $\bm{d}_{A}$ and $\bm{d}_{B}$ of the groups of
  these positions being also complex conjugate,
\begin{equation}
\label{eq:tab:wf114:2}
\bm{d}_{A} = \bm{d}^*_{B}.
\end{equation}     
The matrix {\bf N} defined by Theorem 7 of Ref.\ \cite{theoriewf}
takes the form $ {\bf N} = {\tiny \left(
  \begin{array}{cccc} 
  0 & 1 & 0 & 0\\              
  1 & 0 & 0 & 0\\              
  0 & 0 & 0 & 1\\              
  0 & 0 & 1 & 0\\              
  \end{array}
\right) } $ in band 2 of Mn, yielding the two co-representations
$\bm{d}_{A}$ and $\bm{d}_{B}$ defined by the table
\begin{center}
\begin{tabular}[c]{ccccc}
 & $E$ & $C_{2z}$ & $K\sigma_x$ & $K\sigma_y$\\
\hline
$\bm{d}_{A}$ & 1 & -1 & 1 & -1\\
$\bm{d}_{B}$ & 1 & -1 & -1 & 1\\
\hline\\
\end{tabular}.
\end{center}
Because $\bm{d}_{A}$ and $\bm{d}_{B}$ do not comply with Equation
\gl{eq:tab:wf114:2}, the Wannier functions defined by band 2 of Mn do
not form a magnetic band in antiferromagnetic BaMn$_2$As$_2$ since it
is experimentally proven that the ordered magnetic moments lie at the Mn
atoms. The Wannier functions defined by band 2 of As, on the other
hand, form a magnetic band in BaMn$_2$As$_2$ because the As atoms do
not bear ordered magnetic moments.
\item Each band consists of two or four branches (Definition 2 of Ref.\
  \cite{theoriewf}) depending on the number of the related atoms in
  the unit cell.   
\end{enumerate}
\end{table}

\begin{table}[t]
\caption{
  Character tables of the single-valued irreducible representations of the
  tetragonal space group $P\overline{4} = \Gamma_qS^{1}_{4}$ (81).
  \label{tab:rep81}}
\begin{tabular}[t]{ccccccc}
\multicolumn{7}{c}{$\Gamma (000)$, $M (\frac{1}{2}\frac{1}{2}0)$, $Z
  (00\frac{1}{2})$, $A (\frac{1}{2}\frac{1}{2}\frac{1}{2})$}\\
 & $K$ & $KC_{2a}$ & $E$ & $S^+_{4z}$ & $C_{2z}$ & $S^-_{4z}$\\
\hline
$R_1$ & (a) & (a) & 1 & 1 & 1 & 1\\
$R_2$ & (c) & (a) & 1 & i & -1 & -i\\
$R_3$ & (a) & (a) & 1 & -1 & 1 & -1\\
$R_4$ & (c) & (a) & 1 & -i & -1 & i\\
\hline\\
\end{tabular}\hspace{1cm}
\ \\
\begin{flushleft}
Notes to Table~\ref{tab:rep81}
\end{flushleft}
\begin{enumerate}
\item The notations of the points of symmetry follow Fig. 3.9 
of Ref.~\cite{bc}.
\item Only the points of symmetry invariant under the complete space
  group are listed.
\item The character tables are determined from Table 5.7 in
  Ref.~\protect\cite{bc}.  
\item $K$ still denotes the operator of time inversion. The entries (a) and
  (c) indicate whether the related corepresentations of the magnetic
  groups $P\overline{4} + \{K|000\}P\overline{4}$ and
  $P\overline{4} +
  \{KC_{2a}|000\}P\overline{4}$ follow
  case (a) or case (c) as defined in equations\ (7.3.45) and (7.3.47),
  respectively, of Ref.~\cite{bc} (and determined by equation\
  (7.3.51) of Ref.~\cite{bc}). 
\item 
\label{item:exact}
The entries (a) and (c) for $K$ and $KC_{2a}$ show that the
  representations $R_2$ and $R_4$ at any of the points $\Gamma$,
  $M$, $Z$, or $A$ are possible representations of a {\em stable}
  antiferromagnetic state, see Appendix A of Ref.\ \cite{josla2cuo4}
  or Section III C of Ref.\ \cite{ea}. This is important since
  $M_{81}$~\gl{eq:6} is the {\em exact} group of the magnetic
  structure in BaMn$_2$As$_2$.  
\end{enumerate}
\end{table}

\FloatBarrier

%


\end{document}